\documentclass[letterpaper, 10 pt, journal, twoside]{IEEEtran}
\IEEEoverridecommandlockouts 
\usepackage{url}
\usepackage{bbold}
\usepackage{graphicx,amstext,amsmath,amssymb,color,psfrag}
\usepackage[latin1]{inputenc}
\usepackage{marvosym}
\renewcommand{\EUR}{\unskip~\text{\mvchr{164}}}
\DeclareInputText{128}{\EUR}
\usepackage{paralist}
\usepackage{environ}         
\usepackage{etoolbox}        
\usepackage{graphicx}        

\usepackage[]{times} 
\usepackage{float}
\usepackage{amsfonts}
\usepackage{placeins}
\usepackage{setspace}

\newcommand{\R}{\ensuremath{\mathbb{R}}}

\newcommand{\Rlo}{\ensuremath{\mathbb{R}_{\geq 0}}}

\definecolor{bleucit}{rgb}{0.2,0.4,0.6} 

\definecolor{blue_cv}{rgb}{0.09,0.35,0.78}

\newcommand{\Argmin}{\ensuremath{\text{argmin}\,}}
\DeclareMathOperator*{\argmin}{\Argmin}

\newtheorem{ass}{Assumption}

\newtheorem{prop}{Proposition}

\newtheorem{lem}{Lemma}

\newtheorem{thm}{Theorem}

\usepackage{adjustbox}
\usepackage[lofdepth,lotdepth]{subfig}
\title{On the efficiency of decentralized epidemic management and application to Covid-19 $^*$\thanks{$^*$This work was supported by ANR via the grant NICETWEET, number ANR-20-CE48-0009 and by CNRS via the grant COVEXIT.}}

\author{Olivier Lindamulage De Silva$^1$, Samson Lasaulce$^1$, and Irinel-Constantin Mor\u{a}rescu$^1$\thanks{$^1$Universit\'e de Lorraine, CNRS, CRAN, F-54000 Nancy, France, {\small \tt olivier.lindamulage-de-silva@univ-lorraine.fr}}}

\begin{document}

\maketitle
\thispagestyle{empty}
\pagestyle{empty}
\begin{abstract}
In this paper, we introduce a game that allows one to assess the potential  { loss of efficiency induced by a decentralized control or local management of} a global epidemic.  {Each player typically represents a region or a country which} is assumed to choose its control action to implement a tradeoff between  {socio-economic} aspects and the health aspect. We conduct the Nash equilibrium analysis of this game. Since the analysis is not trivial in general, sufficient conditions for existence and uniqueness are provided. Then we quantify through numerical results the loss induced by decentralization, measured in terms of price of anarchy ($\mathrm{PoA}$) and price of connectedness ($\mathrm{PoC}$). These results allow one to clearly identify scenarios where decentralization is acceptable or not regarding to the retained global efficiency measures.       
\end{abstract}
\vspace*{-1em}
\section{Introduction}

\IEEEPARstart{I}{n} 2020, many governments around the world had to take drastic measures to mitigate the propagation of the SARS-Cov2 virus. Especially over the first half of 2020, similar measures were taken over  {large geographical areas such as countries}. One major drawback from implementing such a (uniform) policy was that there has been a mismatch between the measure severity level and the local situation. Among the consequences of this mismatch we find: avoidable local economic losses, potentially avoidable psychological damages, frustration, and thus a degradation in terms of measure effectiveness. In 2021, the experience acquired on the pandemic shows that allowing regions (e.g., provinces in China, states in the USA, L\"{a}nder in Germany, or regions in France) to locally adjust the decisions may be more suited. This is also true when it comes to vaccination.  {Consequently, different countries adopted different control strategies prioritizing aspects such as education, social welfare, economy, or health. Also within a given country the measures implemented were different for regions depending on the local situation. These measures have been aiming at achieving a certain tradeoff between socio-economic aspects and health aspects. Motivated by this observation, we propose a mathematical model to analyze the effects of decentralization on the epidemic management. In this context, each region or country is a decision-maker. The proposed model is built on existing models such as the networked epidemic models \cite{pastor2015epidemic,stella2020role,kephart1992directed}, \cite[Chapter 9.3]{mesbahi2010graph}}. To assess the potential efficiency loss induced by decentralizing the epidemic management, we consider a mathematical model that is relatively simple while capturing the main effects of interest. {This paper considers a strategic-form game which is built from a networked Susceptible-Infected-Recovered (SIR) compartmental model \cite{magal2016final,mei2017dynamics}. Precisely, we consider a game where each player represents a geographical area which decides social-distancing rules which aim at minimizing a cost; this area may typically correspond to a country or a region of a country but it can also correspond to a metropolis or a group of countries. Each individual cost implements a given trade-off between socio-economic losses and health losses. Indeed, the value of the cost for each region not only depend on its actions but also on the actions of neighboring regions through the network structure and the epidemic dynamics. It is noteworthy that the proposed game model is a static or one-shot game model; a player chooses a control action which is fixed over a given finite time horizon. In practice, this action would need to be updated for each epidemic phase.} Moreover, each region is assumed to have its own virus transmission rate (e.g., depending on local population density, weather conditions, the effectiveness level of the taken measures);  {the propagation among regions occurs with an intensity which is given by the cross transmission rates (e.g., depending on the geographic mobility among regions, the sanitary rules imposed by the corresponding neighbors); the population of each node recovers with a fixed recovery rate (e.g., depending on the capacity and performance of the health system  \cite{hota2020closed}).} 
 {To the best of our knowledge,} the game we introduce in this paper differs from existing works for several reasons. In \cite{elie2020contact}, the players are the individuals but their decision consist in controlling social distancing with others to find a balance between social interactions and the chances of being infected by the virus that spreads through a single region SIR model.\\ 
 {The main differences between the present work and the existing results on networked epidemic games (e.g., \cite{hota2019game2,omic2009protecting,hayel2014complete,trajanovski2015decentralized}) can be summarized as follows: we consider an SIR model while the existing game models are applied to the networked SIS model (Susceptible-Infected-Susceptible);  we control the inter-regions transmission rates while the existing works consider networks described by a binary adjacency matrix; we propose a one-shot game over a finite time horizon unlike the existing works consider infinite time games with constant actions.\\ 
Our key contributions can be summarized as follows:\\
$\bullet$ We formulate a strategic form game applied to a networked SIR model, in which: the interactions are described by a weighted adjacency matrix; a player is a node of the network that tries to minimize its own cost capturing a tradeoff between the socio-economic and health losses; the decision of each player affects the costs of his neighbors in the network; the decision of each player is constant during a finite time horizon (working phase).\\
$\bullet$ We introduce an operating regime called the Weak Interconnection Regime (\textbf{WIR}), that allow us to conduct the Nash equilibrium (NE) of the considered game, by ensuring existence and uniqueness. Furthermore, this analysis ensures the well-posedness of the two equilibrium efficiency measures considered in this work: the Price of Anarchy ($\mathrm{PoA}$) and the Price of Connectedness ($\mathrm{PoC}$).}

This paper is structured as follows. In Sec. II, the considered networked SIR epidemic model is described. The proposed strategic form game model to implement the tradeoff of interest and the chosen measures of global efficiency are provided. In Sec. III, we conduct a complete analysis of the corresponding Nash equilibria (existence, uniqueness). In Sec. IV, we show how the studied game can be exploited numerically for a Covid-19-type-scenario  {and we discuss the effectiveness of the decentralization strategy on the decision-making process.}


 \color{black}
\color{black}

\section{Problem statement}

We consider a set of $K > 1$ interconnected regions (e.g., provinces, states, or cities) that are affected by an epidemic; the region index is denoted by $k \in \mathcal{K}:=\{1,\ldots,K\}$. The epidemic propagation within a region is assumed to follow a SIR model. This section provides both the model that we consider for the epidemic dynamics in the presence of interconnected regions (Sec.~\ref{subsec:epidemic-model}) and the game proposed to model the fact that the epidemic management is decentralized (Sec.~\ref{subsec:game}). 
The proposed game model intends to be simple while capturing a key feature, which is the tradeoff between  {socio-economic} losses, and health aspects. 




\subsection{Epidemic Model} \label{subsec:epidemic-model}

For Region $k \in \mathcal{K}$, we respectively denote by $\beta_{kk}$ and $\gamma_k$ the virus (endogenous) transmission rate and the removal/recovery rate ($\frac{1}{\gamma_k}$ is called the average recovery period). For $k \neq \ell$, the quantity $\beta_{k\ell}$ denotes the transmission rate from Region $\ell$ to Region $k$. The action of Region $k$ on the epidemics is represented by a scalar control action denoted by $u_k \in \mathcal{U}_k$ where $\mathcal{U}_k:= [{U}_k^{\min},{U}_k^{\max}]\subset[0,1)$ is compact. \textbf{The control action $u_k$ is assumed to be constant over the time period of interest (working phase) which is the interval $[0,T]$, $T>0$}. In this paper, we restrict our attention to the study over a single phase; a phase may typically last few weeks. In practice, the action would need to be updated for each phase. This corresponds to considering a blockwise constant management strategy, which is the easiest to implement in practice. We will denote by $u$ the control action profile or vector: $u:=(u_1,\ldots,u_K) \in\mathcal{U}$ where $\mathcal{U}:= \mathcal{U}_1 \times \ldots\times \mathcal{U}_K$ and we will also use the notation $u_{-k}$ to refer to the reduced action profile $u_{-k}:=\left(u_{1},\ldots,u_{k-1},u_{k+1},\ldots,u_{K}\right)$. The fractions of susceptibles, infected, and recovered for Region $k$ are respectively denoted by $s_k(t,u_{k},u_{-k})\in[0,1]$, $i_k(t,u_{k},u_{-k})\in[0,1]$, and $r_k(t,u_{k},u_{-k})\in[0,1]$. With this notation, the continuous-time dynamics for the epidemic in Region $k$ in presence of interconnection is assumed to be given by $\forall T\in\Rlo$, $u\in\mathcal{U}$: 
\begin{equation}\label{eq-sys-tot}
\left\{\begin{array}{l}
\displaystyle{\frac{\partial s_k}{\partial t}(t, {u})}=-s_k(t, {u})\Big[(1-u_k)\sum_{\ell=1}^{K} \beta_{k\ell} i_\ell(t, {u})\Big]\\
\displaystyle{\frac{\partial i_k}{\partial t}(t, {u})}=-\frac{\partial s_k}{\partial t}(t, {u})-\gamma_k i_k(t, {u})\\
\displaystyle{\frac{\partial r_k}{\partial t}(t, {u})}=\gamma_k i_k(t, {u})\\
s_k(t, {u})+i_k(t, {u})+r_k(t, {u})=1,
\end{array}\right.
\end{equation}
where the initial fractions of susceptibles and infected are chosen as $s_k^0>0$ and $i_k^0\geq0$.

  \color{black}For the sake of simplicity we assume that the social distancing rules imposed in Region $k$ (namely, $u_k$) affects uniformly all the infected population of each region in contact with the susceptibles of Region $k$, i.e., $(1-u_k)\beta_{k \ell}$ is  the controlled rate at  which the infected individuals of Region $\ell$ infects the susceptibles in Region $k$. In practice, it would be quite difficult to measure its value, or to assign it a prescribed value. Then, we assume policy makers of each region $k$ would apply a social-distancing rule close enough to the abstract quantity $u_k$. 
 \color{black}


\subsection{Game Model} \label{subsec:game}

Each region is assumed to seek for a tradeoff between the socio-economic losses and the local health impact of the epidemic, induced by the sanitary rules. This amounts to considering a cost function that comprises three terms. Precisely, we assume that a region aims at minimizing the following composite cost:
 {\begin{equation}\label{eq:cost}
J_k(u):=\underbrace{a_k u_{k}+b_k u_{k}^2}_{\textbf{socio-economic losses}} 
    +\underbrace{c_k\Big[s_k^0-s_k(T,u)\Big]}_{\textbf{health losses}},
    \end{equation}}
where $(a_k, b_k, c_k) \in \Rlo^3$  {are constant}.  {The reasoning behind this choice is that social-distancing strategies induce both health and socio-economic losses. In particular, we assume the socio-economic cost is a sum of linear and quadratic terms w.r.t the social distancing rules, as motivated in the related literature of optimal control applied to epidemic that spreads in a single Region (see e.g., \cite{di2020covid}, \cite[Section 2.4]{lasaulce2020efficient}); this assumption seems to be commonly accepted in economic studies, according to \cite[Eq.~8 in Section 2.2.2]{charpentier2020covid}. On the other hand, we consider the health losses to be proportional to the final size of the epidemic after a working phase. In particular, the decision of each node has an impact on its neighbors, through the network structure and the epidemic dynamics.} The strategic form (see e.g., \cite{lasaulce2011game}) of the static game under consideration is therefore given by:
\begin{equation}\label{eq-game}
    \mathcal{G}:=\Big(\mathcal{K}, \Big(\mathcal{U}_{k}\big)_{1\leq k\leq K},\big(J_k\big)_{1\leq k\leq K} \Big),
\end{equation} 
in which the players (nodes of the network) are the regions of a country  {(or simply countries); the action space for Player $k$ is given by $\mathcal{U}_k=[U_{k}^{\min},U_{k}^{\max}]\subset(0,1)$; the individual cost function of Player $k\in\mathcal{K}$ is given by $J_k$ in (\ref{eq:cost}). Region $k\in\mathcal{K}$ expresses its interests by setting the triple $(a_k,b_k,c_k)$, whereas the set of action $\mathcal{U}_k$ is imposed by a social planner (e.g., a country or an international organization, depending on the nature of the player). In the case where players are countries, we assume that the social planner might be a worldwide organization such as the WHO (World Health Organization). In addition, we emphasize that the theoretical results established in this paper hold for a multistage game setup in which the one-shot game is repeated at each stage (for which the parameters are updated) and different constant control actions are applied during it. In this letter we make the choice} not to add a constraint on the region states. For instance, there is no constraint on $i_k(t,u)$ to account e.g., for the number of intensive care units (ICU) in a region.  {To treat the problem in presence of coupling constraints one would need to resort to more advanced notions such as the generalized NE, which is left as an extension of the proposed analysis. Notice that the third term of the cost functions can already be seen as a way of controlling the epidemic and the number of people requiring ICUs.}


\subsection{Efficiency measures}

One of the main objectives of this paper is to assess the potential inefficiencies that might be induced by letting each region choose its control action. A famous and well-used measure of global efficiency is given by the Price of Anarchy (PoA) of a game \cite{papadimitriou2001algorithms}. Before defining the PoA, let us remind the definition of a Nash equilibrium (NE). An action profile $u^{\mathrm{NE}}$ is an NE if: $\forall k, \forall u_k'$, $J_k(u^{\mathrm{NE}}) \leq J_k(u_k', u_{-k}^{\mathrm{NE}})$. The PoA is defined by:
\begin{equation}\label{eq-POA}
     \mathrm{PoA}:=\max\limits_{u\in\mathcal{U}^{\mathrm{NE}}}\displaystyle{\sum_{k=1}^K} J_k(u)\Big/\min\limits_{u\in\mathcal{U}}\displaystyle{\sum_{k=1}^K} J_k(u),
\end{equation} 
where $\mathcal{U}^{\mathrm{NE}}$ is the set of NE of $\mathcal{G}$. The function $\sum_{k=1}^K J_k$ is often referred as the social cost of the game. The PoA thus compares the performance of the worst NE to the performance of the centralized solution. Implicitly, the PoA assumes that the social cost is a relevant metric to measure the global performance.  {In particular, when the $\mathrm{PoA}$ is too high the decentralization strategy will not be effective at the risk of observing selfish behavior from Players.} 
To have a second measure of global efficiency, we also introduce the Price of Connectedness (PoC), which is defined as follows:
\begin{equation}
     \mathrm{PoC}:=\max\limits_{u\in\mathcal{U}^{\mathrm{NE}}}\displaystyle{\sum_{k=1}^K} J_k(u)\Big/ \displaystyle{\sum_{k=1}^K}
     \min\limits_{u_k\in\mathcal{U}_k}
     \widetilde{J}_k(u_k),\label{eq-PoC}
\end{equation}
where $\widetilde{J}_k(u_k)$ is the cost that Region $k$ would obtain if  {they do not consider the influence of the network i.e.,} the crossing transmission rates $\beta_{k\ell}$, $k\neq \ell$, would be vanishing in (\ref{eq-sys-tot}). This therefore corresponds to the performance that Region $k$ would expect to obtain by neglecting the interactions with the other regions while these actually exist, hence the term PoC.  {Such as for the other efficiency measure, we consider that when the $\mathrm{PoC}$ is too high Regions should take into account the network structure before taking a decision.}

\section{Nash equilibrium analysis}
Since one of our main objectives is to measure efficiency at NE  {through the $\mathrm{PoA}$ and $\mathrm{PoC}$ in \eqref{eq-POA}-\eqref{eq-PoC}, it is necessary to conduct the complete equilibrium analysis of the NE. This analysis} includes the study of the existence and uniqueness of the NE. 


\subsection{Existence} 
In this section, we state our main result concerning the existence of a pure NE. Notice that the existence of a mixed NE is ensured by the  continuity of the cost functions $J_k$, $k \in \mathcal{K}$ (see \cite{lasaulce2011game}), but it is of no practical interest in our setting. The existence of a pure NE is strongly related to the geometrical properties of the cost functions $J_k$, $k \in \mathcal{K}$, such as  {the quasi-convexity} properties. Since the dependency of the  {third term} of $J_k$ on $u_k$ is not explicit,  {the quasi-convexity} analysis of $J_k$ appears to be a non-trivial problem. This is the reason why we define  {a} working regime in which it is possible to prove that $J_k$ is  {quasi-convex} w.r.t. $u_k$.

 \color{black}

\textit{Weak  {Interconnection} regime (\textbf{WIR})}: The game $\mathcal{G}$ is said to be in the \textbf{WIR}, if $\forall (k,\ell)\in\mathcal{K}^2:\ \ell \neq k$ there exists $\nu_{\beta,k}>0$ such that $\beta_{k \ell} \leq \nu_{\beta,k}$ and $J_k$ is quasi-convex w.r.t. $u_k$ on $\mathcal{U}_k$ (i.e, $\forall u_{-k}\in\mathcal{U}_{-k}$, $\forall \lambda\in\R$, the lower level set $\mathcal{L}^k(u_{-k},\lambda):=\left\{u_k\in\mathcal{U}_k:\ J_k(u_k,u_{-k})\leq \lambda\right\}$ is convex). 

 \color{black}The motivation behind the definition of the \textbf{WIR} is given by the following result. 
\begin{prop}\label{prop-existenceNE}
In the \textbf{WIR} the game $\mathcal{G}$ has at least one pure NE.
\end{prop}
\noindent\emph{\textbf{Proof:}}
See Appendix\ref{Appendix-Thm-ExistenceNE}.\hfill$\blacksquare$

An important practical question would be: "When is the game in the \textbf{WIR}?". To answer this technical question, let us introduce the following working assumption.
\begin{ass} Let $\forall (k,\ell) \in \mathcal{K}^2, \rho_{k\ell}:= \beta_{k\ell}  / \gamma_{\ell}$.\\
\noindent Condition (i): The matrix $\widehat{\boldsymbol{B}}$ whose entries are given by: $\widehat{\boldsymbol{B}}_{k,\ell}=\beta_{k \ell}$, is non-singular.\\
\noindent Condition (ii): $\forall k$, $\forall u$, $\forall T\in\mathcal{T}$ one has that $s_k(t,u)>0$.\\
\noindent Condition (iii): $\mathcal{T}=\Rlo$ where\\
\hspace*{-0.3em}$\mathcal{T}\hspace{-0.2em}:=\left\{t\in\Rlo: \forall k\text{, } \forall u\text{, }(1-u_k)s_k(t,u)\leq\frac{1}{\sum_{\ell=1}^K \rho_{k \ell}}\right\}.$
\label{ass1}
\end{ass}
 \color{black}Condition (i) is ensured when $\widehat{\boldsymbol{B}}$ is strictly diagonally dominant (which is often the case in practice because intra-regions interactions are much stronger than inter-regions ones); Condition (ii) is trivially satisfied as far as the epidemic does not affect the entire population; Condition (iii) is needed to characterize a bound for the inter-regions interactions i.e., to quantitatively describe the \textbf{WIR} with $\nu_{\beta,k}:=\left(\frac{\min\limits_{\ell\in\mathcal{K}}\gamma_\ell}{(1-U_k^{\min})s_k^0}-\beta_{k k}\right)\Big/4.$ 
In what follows, we propose to exhibit a sufficient condition such that the game is in the \textbf{WIR}. \color{black}To establish the corresponding result, a few notations are in order. Let $T\in\mathcal{T}$, $u\in\mathcal{U}$ and $s(T,u)=(s_1(T,u),\ldots,s_K(T,u))^\top,\ i(T,u)=(i_1(T,u),\ldots,i_K(T,u))^\top$, $r=(r_1(T,u),\ldots,r_K(T,u))^\top$. To be able to express the derivative of $s_k$ w.r.t. $u_k$ and exploit the implicit function theorem, let us introduce the two square matrices $\boldsymbol{B}:=\mathrm{diag}(1-u)\widehat{\boldsymbol{B}}$ and $\boldsymbol{\Gamma}:=\mathrm{diag}(\gamma)$, where $\gamma:=(\gamma_1,\ldots,\gamma_K)$.
The reformulated system (\ref{eq-sys-tot}) in a collective dynamics form: $\forall t\in [0,T]$,
\begin{equation}\label{eq-Matrixdyn}
\left\{\begin{array}{l}
    \displaystyle\frac{\partial s}{\partial t}(t, {u})=- \mathrm{ \mathrm{diag}}(s(t, {u}))\boldsymbol{B}i(t, {u})\\
    \displaystyle\frac{\partial i}{\partial t}(t, {u})= \mathrm{ \mathrm{diag}}(s(t, {u}))\boldsymbol{B}i(t, {u})-\boldsymbol{\Gamma} i(t, {u})\\
    \displaystyle\frac{\partial r}{\partial t}(t, {u})=\boldsymbol{\Gamma} i(t, {u}).
\end{array}\right. 
\end{equation} Using (\ref{eq-Matrixdyn}) and \cite[Section 2]{magal2016final}, one can write the following identity:
\begin{equation}
\displaystyle\frac{\mathrm{d}}{\mathrm{d}t}[\boldsymbol{B\Gamma^{-1}}\left(s(t,u)+i(t,u)\right)-\ln(s(t,u))]=0.\label{eq-dintegr}
\end{equation}
Therefore, by integrating \eqref{eq-dintegr} on $[0,T]$, one has that
\begin{equation*}
\begin{aligned}
&\boldsymbol{B\Gamma^{-1}}(s(T,u)+i(T,u)-x^0)=\ln(s(T,u))-\ln(s^0),\\
\end{aligned}
\end{equation*} where $s^0=s(0,\cdot)$, $i^0=i(0,\cdot)$ and $x^0=s^0+i^0$. Let $F:\mathcal{U}\times(0,1]^{2K}\to\R^K$ such that, for any $k\in\mathcal{K}$, the $k^\text{th}$-component of $F$ is given by $F_k:\mathcal{U}\times (0,1]^{2K}\to\R$: 
\[
F_k(u,s,i)=\displaystyle (1-u_k)\sum_{\ell=1}^K\rho_{k\ell}\left(s_\ell+i_\ell-x_\ell^0\right)+\ln\left(\frac{s_k^0}{s_k}\right).\]
 \color{black}We define the set of non-monotonic players as $\mathcal{K}_{\mathrm{NM}}:=\left\{k\in\mathcal{K}: J_k\text{ is not monotone w.r.t. }u_k\right\}$, (i.e., $k\in\mathcal{K}_{\mathrm{NM}}$ if the assigned weights of socio-economic and health losses are such as $J_k$ is non-monotone w.r.t. $u_k$). 

Now that we have introduced all the notations needed to establish the main result of this letter, let us exhibit the following key Lemma that provides, $\forall k\in\mathcal{K}$, a lower-bound on the derivative of $s_k$ w.r.t. $u_k$.
\begin{lem} 
Under Assumption \ref{ass1}, $\forall T\in\mathcal{T},\ \forall u\in\mathcal{U}$ and $\forall (k,\ell)\in\mathcal{K}^2$ one has $\frac{\partial s_k}{\partial u_{\ell}}(T,u)\geq0 $ and\\ $
\hspace*{2em} \displaystyle\frac{\partial s_k}{\partial u_{k}}(T,u)\geq  \frac{ s_k(T,u)\ln\left(\displaystyle\frac{s_k(T,u)}{s_k^0}\right)}{\displaystyle (1-u_k)\left[ (1-u_k)\rho_{k k}s_k(T,u)-1\right]}.$ 
\label{prop-WIR}
\label{lem1}
\end{lem}

\noindent\emph{\textbf{Proof:}}
See Appendix\ref{Appendix-ProofLem1}.\hfill$\blacksquare$

The following Theorem establishes the main result of this paper, by ensuring that the game $\mathcal{G}$ is in the \textbf{WIR}.
 {\begin{thm}\label{Thm1} Let $T\in\mathcal{T}$. Suppose Assumption \ref{ass1} holds and the less restrictive action profile $u_{\min} = (U_1^{\min}, ..., U_K^{\min})\in [0,1)^{K}$ verifies that, $\forall k\in\mathcal{K}_{\mathrm{NM}}$,\\ $\hspace*{5em}(1-U_k^{\min})s_k(T,u_{\min})\geq 1\big/(2\rho_{k k})$.\\ Then, the game $\mathcal{G}$ is in the \textbf{WIR}. 
\end{thm}}
\noindent\emph{\textbf{Proof:}}
See Appendix\ref{Appendix-ProofThm1}.\hfill$\blacksquare$

\color{black}
The additional condition we introduce means that if the epidemics are sufficiently controlled,  {then the game $\mathcal{G}$ is a quasi-convex game that ensures the existence of a pure NE, according to Proposition \ref{prop-existenceNE}.} In practice, that would mean that the social planner would need to track the regions at least partially (e.g., by imposing some minimum epidemic management measures).

 
\subsection{Uniqueness}

 {In practice having the uniqueness of the NE may be a useful feature
for a government (when players are the regions) or for an international organization (when players are countries). It is typically convenient to be able to predict the outcome of the game. If the game models the interactive situation sufficiently well, an NE can be effectively observed. If there is only one NE, the situation becomes predictable, which is not the case in the presence of multiple equilibria.} It is known that uniqueness typically requires additional conditions  (\cite{lasaulce2011game}). The following result establish  {the uniqueness property of the NE, and the convergence of the sequential best-response dynamics.}
 \color{black}
\begin{thm}\label{Thm2} Suppose that $\forall k\in\mathcal{K}$,\[
\displaystyle\frac{\partial^2 J_k}{\partial {u_{k}}^2}(u)>\sum_{\ell=1,\ell\neq k}^K \left\vert\frac{\partial^2 J_k}{\partial {u_{k}}\partial u_\ell}(u)\right\vert.\] Then $\mathcal{G}$ has a unique NE, and the sequential best-response dynamics converges to this equilibrium. 
\end{thm}
\noindent\emph{\textbf{Proof:}}
See Appendix\ref{Appendix-ProofThm2}.\hfill$\blacksquare$

We should note that, if the conditions of Theorem \ref{Thm1} hold for all $k\in\mathcal{K}$, then \color{black}$J_k$ is strictly convex w.r.t. $u_k$ that is, $ \frac{\partial^2 J_k}{\partial {u_{k}}^2}(u) > 0$. Here, the additional condition of Theorem \ref{Thm2} requires that the dependency of the  {second} derivative of $J_k$ w.r.t. the control actions of the other regions is sufficiently small. The latter is both useful to predict the epidemic tendency when its management is decentralized and to compute the NE  {(so the $\mathrm{PoA}$ and $\mathrm{PoC}$)}.

\textit{Remark.}  { If the game is not in the \textbf{WIR} but $\mathcal{K}_{\mathrm{NM}}=\emptyset$, the costs $J_k$ are all individually quasi-convex and the existence of a pure NE is ensured. Moreover, there is a unique pure NE which lies at the extreme of the interval $\mathcal{U}$, in particular whatever the values of $\beta_{k\ell}$. \label{divergentintersts_use_in_ex}}

\section{Numerical performance analysis}

\vspace{-0.2em}The goal of this section is to quantify the $\mathrm{PoA}$ and $\mathrm{PoC}$ numerically for a Covid-19-type scenario. The proposed methodology can be applied to other epidemic scenarios where multiple regions are involved. Motivated by a scenario which has been studied by the French government in May 2020. We assume that France is divided in $K=5$ regions and we propose to observe the influence of the inter-region virus transmission rates $\beta_{k \ell}$ on the $\mathrm{PoA}$ and $\mathrm{PoC}$. To choose the epidemic's parameters, we have exploited the studies on Covid-19 that have been conducted in \cite{lasaulce2020efficient,casella2020can,salje2020estimating}.  \color{black}We assume that: Regions $k\in\{1,2\}$ have selected the weight $a_k, b_k$ and $c_k$ such that only the socio-economic losses matter; Regions $k\in\{3,4,5\}$ weighted the weights of each of the losses such that $\mathcal{K}_{\mathrm{NM}}=\{3,4,5\}$; see the Table \ref{tab1}.
\begin{table}[H]\centering
\begin{tabular}{|c|c|c|c|c|c|c|c|c|c|}
\hline
    $k$ & $\gamma_k$ & $\beta_{k k}$ & $s_k^0$ & $i_k^0$& $a_k$ & $b_k$ & $c_k$\\
    \hline
    $1$ & $0.15 $&  $3\gamma_1$ & $0.8$ & $0.2$& $2$ & $0$& $0$ \\
    \hline
        $2$ & $0.15 $& $2\gamma_2$& $0.9 $& $0.1$&$0.5$ & $0$& $0$ \\
    \hline
        $3$ & $0.15$ &  $1.5\gamma_3$ & $0.9$ & $0.005$&$5$&$2$&$50$ \\
    \hline
        $4$ & $0.15$ &  $1.2\gamma_4$ & $0.9$ & $0.002$&$2$&$5$&$70$\\
    \hline
        $5$ & $0.15$ &  $1\gamma_5$ & $0.9$ & $0.001$&$3$&$5$&$70$\\
    \hline
\end{tabular}
 \caption{ \centering\label{tab1} \small{Epidemic and Game parameters}} 
  \end{table}

\begin{figure}[h!]
\vspace*{-1.1em}
 \centering
\includegraphics[width=\linewidth]{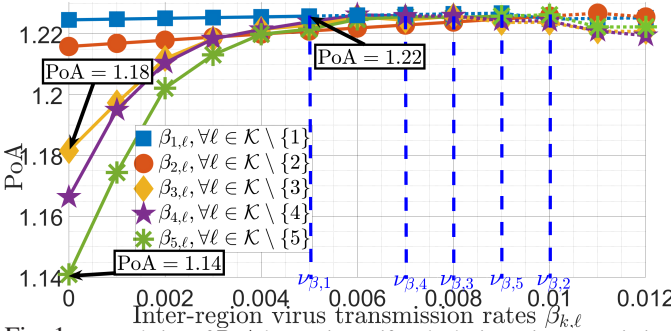} \vspace{-2em}
\caption{\label{fig1}\footnotesize{Interpolation of $\mathrm{PoA}$ by varying uniformly the incoming transmission rates of each Region $k$. The dotted curves do not fit into our theoretical setup.}}
\end{figure}

\begin{figure}[h!]
\centering
  \vspace{-2em}
\includegraphics[width=\linewidth]{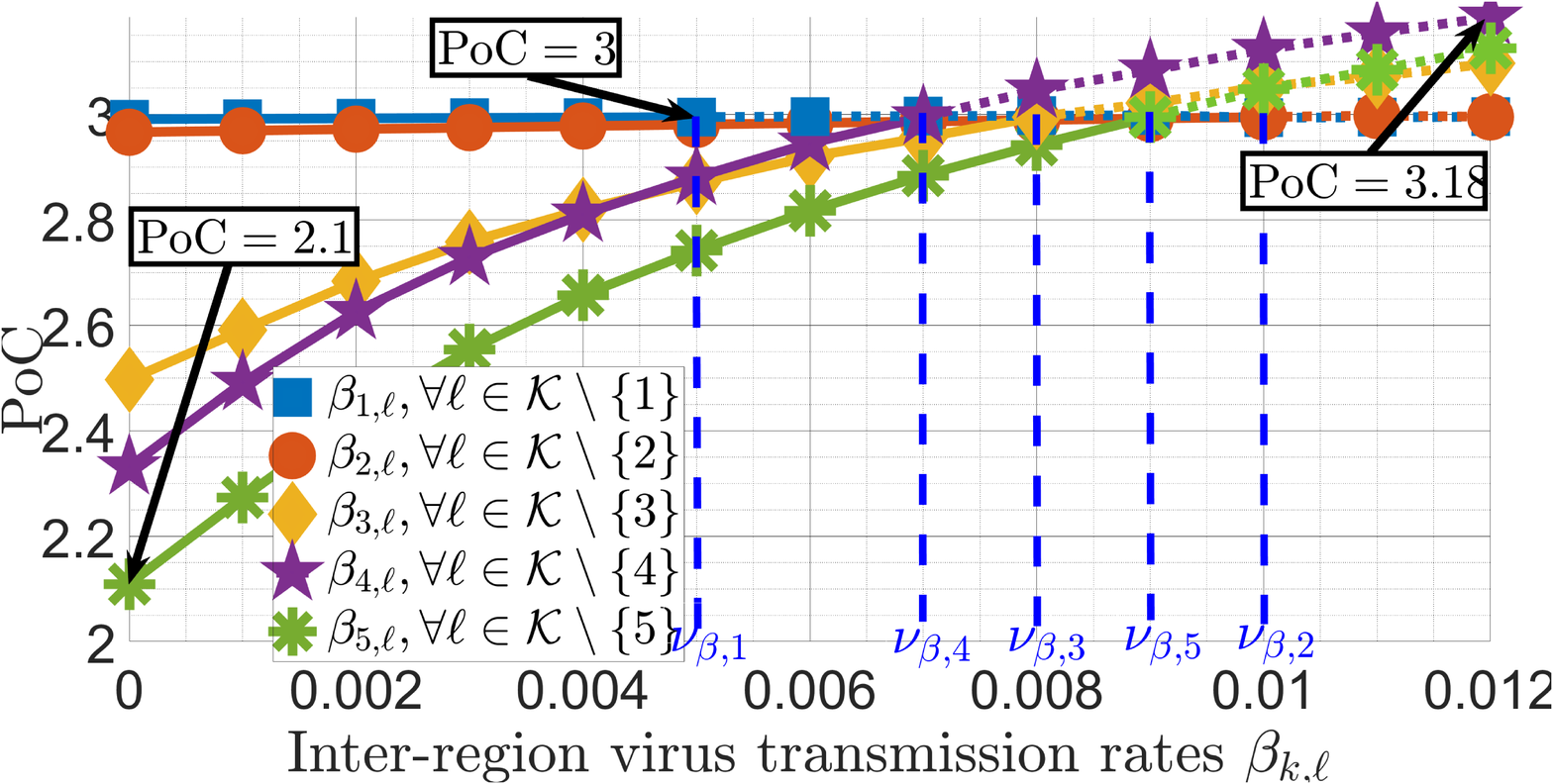} \vspace*{-2em}
\caption{\label{fig2}\footnotesize{Interpolation of $\mathrm{PoC}$ by varying uniformly the incoming transmission rates of each Region $k$. The dotted curves do not fit into our theoretical setup.}}
\end{figure}

\color{black}The time horizon of the considered epidemic phase is set to $T= 30\ \mathrm{days}$ \cite{covidstat,cauchemez2020sortie,santepublic}. The coupled SIR model is implemented by using the \textsf{Matlab ODE45} solver with the Runge-Kutta scheme.  \color{black}The action space is chosen by the social planner such as: $\forall k\in\mathcal{K}$, $U_k^{\max}=0.9$, $u_{\min}=\left(0.6,0.51,0.35,0.2,0.1\right)$ and $\mathcal{U}_k=\{U_{k}^{\min},(U_{k}^{\max}-U_{k}^{\min})\cdot0.1,\ldots,U_{k}^{\max}\}$. In view of the Table \ref{tab1}, the Theorem \ref{Thm1} holds, when the inter-region virus transmission rates $\beta_{k \ell}$ are lower than the constant threshold $\nu_{\beta,k}=\frac{1}{4}\cdot(\frac{\gamma_k}{(1-U_k^{\min})s_k^0}-\beta_{k k})$, which is reasonable in view of the situation in France provided by the National Institute of Statistics and Economic Studies (INSEE) in \cite[Table 6-8]{guan2020transport}. By applying an exhaustive search to find the NE and the social optimal, we show in Fig.~\ref{fig1} and \ref{fig2} the interpolation of the $\mathrm{PoA}$, $\mathrm{PoC}$ w.r.t $\beta_{k \ell},\ \forall k\neq \ell$. Each curve corresponds to a scenario where all incoming transmission rates from a given region vary uniformly (i.e.,$ \forall \ell \neq k,\ \beta_{k\ell} \in\{0,1\cdot10^{-3},\ldots,1.2\cdot 10^{-2}\}$), whereas the other transmission rates are fixed at the threshold value $\nu_{\beta,k}$. We observe that the $\mathrm{PoA}$ can be as large as 1.2 for crossing rates greater than 0.2$\%$. Therefore, the outcome in this case is that the social planner should not decentralize the decision making. We emphasize that, when $\beta_{k \ell}\geq \nu_{\beta,k}$ the simulation does not fit into our theoretical setup. The $\mathrm{PoC}$ measures the impact of ignoring the connection with other regions is even larger and reaches values as large as 3, which shows that a region has a strong interest in accounting for the crossing rates to manage the epidemic locally.

\begin{figure}[h!]
\vspace*{-2em}
\centering
\includegraphics[width=\linewidth]{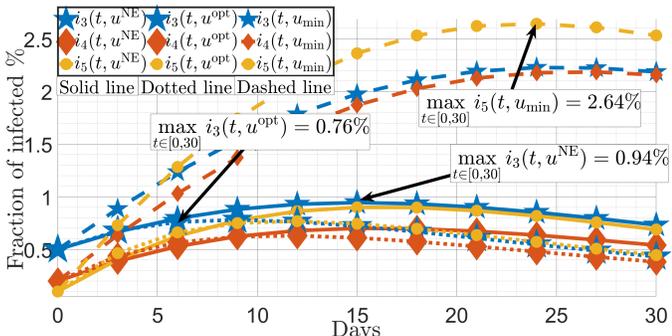} \vspace*{-2em}
\caption{\label{fig3}\footnotesize{Interpolation of infected proportions in each Regions $k\in\{3,4,5\}$. $u^{\mathrm{NE}}$= Nash equilibrium strategy; $u^{\mathrm{opt}}$= optimal centralized strategy; $u_{\min}$= less restrictive policy. }}
\vspace*{-1em}
\end{figure}

 In view of the weights $a_k,b_k,c_k$ given in the Table \ref{tab1}, a natural question should be raised: ``How is the epidemic spreading in the regions $k\in\{3,4,5\}$?'' Fig.~\ref{fig3} shows the evolution over the time of $i_k$, for $k\in\{3,4,5\}$, when different strategy is considered and $\forall k\neq \ell,\ \beta_{k \ell}=\nu_{\beta,k}$. Quantitatively we observe that: when either the NE or optimal strategy is applied, the maximum proportion of infected in Regions $k\in\{3,4,5\}$ is less that $0.94\%$, i.e. if the population sizes in Regions $k\in\{3,4,5\}$ are similar to the region ``\^Ile-de-France'', then the infected proportions are upper-bounded by $112\ 800$ cases, when policy-makers apply either NE or centralized strategy.

 \color{black}

\FloatBarrier

 
\section{Conclusion}

The conducted Nash equilibrium analysis of the proposed game largely relies on the individual  {quasi-convexity} of the cost function of a region. Because one cannot express the state of the fraction of "susceptibles" as a function of the control actions, this analysis turns out to be non-trivial. We exhibit a regime in terms of coupling degree among the regions in which existence is guaranteed; this regime appears to be non-limiting for real scenarios. The numerical analysis allows one to  {clearly quantify what is lost when regions or countries decide by themselves the way to manage the epidemic locally, without coordination. The proposed approach might be improved e.g., by integrating coupled constraints, by investigating a dynamical game formulation of the problem, or by performing a deeper numerical analysis on the impact of the graph on the price of anarchy and the price of connectedness.}   



\appendices
 \section*{Appendix}


 \subsection{Proof of Proposition \ref{prop-existenceNE}}\label{Appendix-Thm-ExistenceNE}
 \color{black}Since the action space of each player $\mathcal{U}_k$ is a convex, compact and non-empty set; the costs $J_k$ are jointly continuous that is continuous w.r.t. the action profile $u\in\mathcal{U}$; the costs $J_k$ are quasi-convex w.r.t. $u_k$ on $\mathcal{U}_k$. Then, the game $\mathcal{G}$ is a quasi-convex game. By Debreu-Fan-Glicksberg theorem for quasi-convex games \cite[Theorem 50]{lasaulce2011game}, the existence of a pure NE is guaranteed.
\color{black}

 \subsection{Proof of Lemma \ref{lem1}}
 \label{Appendix-ProofLem1}
Let $k\in\mathcal{K}$, $u\in\mathcal{U}$, $T\in\mathcal{T}$ and, ${X}:=\Big({u}, {s}(T,u),$ ${i}(T,u)\Big)\in\mathcal{U}\times(0,1]^{K}\times(0,1]^K$ such that $F( {X})=0$. In what follows, we denote by:
\[\begin{array}{l}\boldsymbol{D}:=\left( \mathrm{diag}(s(T,u))^{-1}- \mathrm{diag}(\boldsymbol{B\Gamma^{-1}})\right)^{-1},\\
\boldsymbol{\overline{B}}:=\boldsymbol{B}- \mathrm{diag}(\boldsymbol{B}).\\ 
\end{array}\]
In view of the expression of $F$, we have that:\\
 {$\displaystyle\frac{\partial F}{\partial s}(X)=-\boldsymbol{D^{-1}}\left(\boldsymbol{I}_K-\boldsymbol{D\overline{B}\Gamma^{-1}}\right)$},\\
$\displaystyle\frac{\partial F}{\partial u}(X)=-\mathrm{diag}(\underline{1}-u)^{-1}\mathrm{diag}\left(\ln\left(s(T,u)\right)-\ln\left(s^0\right)\right)$.\\ 
\noindent According to Condition (iii) in Assumption \ref{ass1}, we derive that, {
\begin{equation*}
    \hspace{-0.5em}\left\Vert \boldsymbol{D\overline{B}\Gamma^{-1}}\right\Vert_\infty\hspace{-0.5em}= 
        \displaystyle\max\limits_{k\in\mathcal{K}}\sum_{\ell=1,\ell\neq k}^K\left\vert\frac{(1- {u}_k) \rho_{k \ell}}{\displaystyle {s}_k(T,u)^{-1}-(1- {u}_k)\rho_{k k}}\right\vert<1.
\end{equation*}}
Therefore, the Neumann series converges and,\\ 
 {\hspace{4.5em}$\big(\boldsymbol{I}_K-\boldsymbol{D\overline{B}\Gamma^{-1}}\big)^{-1}=\sum_{k=0}^{+\infty}\left(\boldsymbol{D\overline{B}\Gamma^{-1}}\right)^k.$}\\
 According to the implicit function theorem, it follows that,  {$
\hspace{5em}\displaystyle\frac{\partial s}{\partial u}(T,u)=-\left[{\displaystyle\frac{\partial F}{\partial s}}(X)\right]^{-1}\displaystyle\frac{\partial F}{\partial u}(X).$}\\
We denote by $\widehat{\frac{\partial {s}}{\partial u}}(T,u)$ the approximation of $\frac{\partial {s}}{\partial u}(T,u)$ at the first order of the Neumann series, such that,
 {
$\forall k,\ell,\ \widehat{\frac{\partial {s}}{\partial u}}(T,u):=\big(\boldsymbol{I}_K+\boldsymbol{D\overline{B}\Gamma^{-1}}\big)\boldsymbol{D}\frac{\partial F}{\partial u}(X)$}. Therefore, $\forall k,\ell,\ \frac{\partial s_k}{\partial u_\ell}(T,u)\geq  \widehat{\frac{\partial s_k}{\partial u_\ell}}(T,u)\geq0$, since Condition (iii) of Assumption \ref{ass1} holds.  {The lower bound of $\frac{\partial s_k}{\partial u_k}(T,u)$ given in Lemma \ref{lem1} corresponds to $\widehat{\frac{\partial s_k}{\partial u_k}}(T,u)$.}
\color{black}

\vspace{-2mm}
\subsection{Proof of Theorem \ref{Thm1}}
\label{Appendix-ProofThm1}
 \color{black}
The goal of this proof is to ensure that $\forall k\in\mathcal{K}$, $J_k$ is quasi-convex w.r.t $u_k\in\mathcal{U}_k$. We know that, $\forall k\in\mathcal{K}\setminus\mathcal{K}_{\mathrm{NM}}$, $J_k$ quasi-convexity property holds. In what follows, we are interest in to show the convexity of costs $J_k$ for Players $k\in\mathcal{K}_{\mathrm{NM}}$. Therefore, we propose to analyze in a first step the convexity of $i_k$ w.r.t $u_k$, which allows us to discuss about the concavity of $s_k$ w.r.t $u_k$ for $k\in\mathcal{K}_{\mathrm{NM}}$.

Let $k\in\mathcal{K}_{\mathrm{NM}}$, $u\in\mathcal{U}$, $T\in\mathcal{T}$ and ${X}:=\Big({u}, {s}(T,u),{i}(T,u)\Big)\in\mathcal{U}\times(0,1]^{K}\times(0,1]^K$ such that $F({X})=0$. By following the same reasoning as in Lemma \ref{lem1}, we apply the implicit function theorem to the function $F:\mathcal{U}\times (0,1]^K\times (0,1]^K\to\R^K$ with: $\displaystyle\frac{\partial F}{\partial i}(X)=\mathrm{diag}(1-u)\widehat{\boldsymbol{B}}\Gamma$.
Hence, we derive that
\[\displaystyle\frac{\partial i_k}{\partial u_k}(T,u)=\displaystyle\left[\frac{\partial F}{\partial i}^{-1}\hspace{-1em}(X)\frac{\partial F}{\partial u}(X)\right]_{k,k}=\frac{\gamma_k b_{k k}^{\mathrm{inv}}\ln\left(\frac{s_k(T,u)}{s_k^0}\right)}{(1-u_k)^2},\]
where $b_{k k}^\mathrm{inv}$ is the $(k,k)^\text{th}$ element of $\widehat{\boldsymbol{B}}^{-1}$. Let $u_{-k}\in\mathcal{U}_{-k}$, $\lambda\in\R$, $\displaystyle (\underline{u}_k,\overline{u}_k)\in\Big\{u_k\in\mathcal{U}:$ $\displaystyle\frac{\partial i_k}{\partial u_k}(T,u)\leq \lambda\Big\}$ such that $\underline{u}_k\leq \overline{u}_k$. Given that: (i) $\forall \alpha\in[0,1]$, $\underline{u}_k\leq\alpha\underline{u}_k+(1-\alpha)\overline{u}_k\leq\overline{u}_k$; (ii) $s_k$ is increasing w.r.t. $u_k$; we derive the quasi-convexity of $\frac{\partial i_k}{\partial u_k}$ w.r.t $u_k$, since the following holds.
\[\vspace{-0.2em}\hspace{-0.3em}\displaystyle\frac{\partial i_k}{\partial u_k}(T,\alpha\underline{u}_k+(1-\alpha)\overline{u}_k,u_{-k})\leq\frac{\gamma_k b_{k k}^{\mathrm{inv}}\ln\left(\frac{s_k(T,\overline{u}_k,u_{-k})}{s_k^0}\right)}{(1-\underline{u}_k)^2}\hspace{-0.2em}\leq\hspace{-0.2em}\lambda.\]

\indent Let us write the second derivative of $i_k$ w.r.t. $u_k$, $\displaystyle\frac{\partial^2 i_k}{\partial u_k^2}(T,u)$ $=\frac{\gamma_k b_{k k}^{\mathrm{inv}}\left(\frac{\partial s_k}{\partial u_k}(T,u)(1-u_k)+2s_k(T,u)\ln\left(\frac{s_k(T,u)}{s_k^0}\right)\right)}{(1-u_k)^3 s_k(T,u)}.$
By combining with the lower-bound of $\frac{\partial s_k}{\partial u_k}$ given in Lemma \ref{lem1}, we derive that $\frac{\partial^2 i_k}{\partial u_k^2}(T,u)\geq\frac{\gamma_k b_{k k}^{\mathrm{inv}} \ln\left(\frac{s_k(T,u)}{s_k^0}\right)}{(1-u_k)^3}G_k(u)$ where $G_k(u):=\left(\frac{-\gamma_k}{\gamma_k-s_k(T,u)(1-u_k)\beta_{k k}}+2\right).$ In view of the condition given in Theorem \ref{Thm1}, it follows that $G_k(u_{\min})\leq 0$ then $\frac{\partial^2 i_k}{\partial u_k^2}(T,u_{\min})\geq 0$. Since $\frac{\partial i_k}{\partial u_k}$ is quasi-convex w.r.t. $u_k$, then $\forall T\in\mathcal{T}$ and $\forall u\in\mathcal{U},\ \frac{\partial^2 i_k}{\partial u_k^2}(T,u)\geq 0$. Since, $\forall k,\ r_k(T,u)=\int_{0}^T \gamma_k i_k(t,u) dt$, it follows from the Leibniz's rule for differentiation under the integral sign that $\frac{\partial^2 r_k}{\partial u_k^2}(T,u)=\int_{0}^T \gamma_k\frac{\partial^2 i_k}{\partial u_k^2}(t,u)dt\geq0$. Hence, $\forall T\in\mathcal{T},\ \forall u\in\mathcal{U}$, $\frac{\partial^2 s_k}{\partial u_k^2}(T,u)\leq 0$, since $s_k=-i_k-r_k$.

To conclude this proof, $\forall k\in\mathcal{K}$, $J_k$ is quasi-convex then by definition the game $\mathcal{G}$ is in the \textbf{WIR}.\hfill
\color{black}
\vspace{-2mm}
\subsection{Proof of Theorem \ref{Thm2}}
\label{Appendix-ProofThm2}

According to \cite[Section 2.5-2.6]{vives1999oligopoly}, a sufficient condition to ensure the contraction of the Best-response mapping given by, $
    \mathrm{BR}(\cdot)=\left(\argmin\limits_{u\in\mathcal{U}_1} J_1(u,\cdot),\ldots,\argmin\limits_{u\in\mathcal{U}_K} J_K(u,\cdot)\right)
$ is to verify the strict diagonal dominance condition, which yields that: $\left[\nabla^2 J \right]_{1\leq k,\ell\leq K}=\left[\frac{\partial^2 J_k}{\partial u_k\partial u_\ell}\right]_{1\leq k,\ell\leq K}\hspace{-3.5em}>0\Rightarrow\nabla^2 J+ \nabla^2 {J}^\top>0.$ Hence, according to \cite[Theorem 2 and Theorem 6] {rosen1965existence}, the diagonally strictly convexity (DSC) condition is verified that ensures the uniqueness of the NE. Moreover, in view of \cite[Section 2.5]{vives1999oligopoly}, the sequential best-response algorithm converges to the unique Nash equilibrium of the game $\mathcal{G}$.

%


\color{black}

\bibliographystyle{unsrt}

\end{document}